\documentclass[letterpaper, 10 pt, conference]{ieeeconf}  

\IEEEoverridecommandlockouts                              
\usepackage{epsfig} 
\usepackage{amsmath,amsfonts,amssymb}
\usepackage{algorithm,algorithmicx,algpseudocode}
\usepackage[utf8]{inputenc} 
\usepackage[T1]{fontenc}
\usepackage{url}
\usepackage{hyperref}
\usepackage{tabularx}
\usepackage{booktabs}
\usepackage{cleveref}

\usepackage{xcolor}

\usepackage{graphicx,float}
\graphicspath{{figures/}}

\usepackage{todonotes}

\def\cN{\mathcal{N}}
\def\cF{\mathcal{F}}

\def\cJ{\mathcal{J}}
\def\-{\text{-}}
\def\+{\text{+}}

\newcommand\given[1][]{\:#1\vert\:}

\newtheorem{theorem}{Theorem}[section]
\newtheorem{lemma}[theorem]{Lemma}

\title{\LARGE \bf Information-seeking polynomial NARX model-predictive control\\ through expected free energy minimization}

\author{Wouter M. Kouw%
\thanks{WM Kouw is with the Bayesian Intelligent Autonomous Systems laboratory, Department of Electrical Engineering, TU Eindhoven, Postal box 513, 5600 MB Eindhoven, Netherlands. {\tt\small @: w.m.kouw@tue.nl}}%
}

\begin{document}

\maketitle
\thispagestyle{empty}
\pagestyle{empty}

\begin{abstract}
We propose an adaptive model-predictive controller that balances driving the system to a goal state and seeking system observations that are informative with respect to the parameters of a nonlinear autoregressive exogenous model. The controller's objective function is derived from an expected free energy functional and contains information-theoretic terms expressing uncertainty over model parameters and output predictions. Experiments illustrate how parameter uncertainty affects the control objective and evaluate the proposed controller for a pendulum swing-up task.
%
\end{abstract}
\section{Introduction}
Nonlinear autoregressive exogenous (NARX) models are useful black-box descriptions of dynamical systems
\cite{billings2013nonlinear}. However, a model-predictive controller that relies on a NARX model that has yet to accurately capture dynamics can behave erratically, which may lead to damage during operation \cite{grune2017nonlinear}. To avoid such risks, in practice, model parameters are often first estimated using a designed control signal (i.e., offline system identification). However, this is time-consuming. Here we present a model-predictive controller (also referred to as an agent) that seeks control inputs for rapid online system identification, when its parameter uncertainty and predictive variance are high. When these are low, it seeks control inputs that lead to the goal output.

The proposed agent utilizes variational Bayesian inference, a form of statistical inference that optimizes an analytic approximation to posterior distributions \cite{mackay2003information}. 
Specifically, it minimizes an expected free energy (EFE) functional, a quantity proposed in the neuroscience community that describes perception and action in terms of information-theoretic quantities (also known as active inference) \cite{friston2015active}. Free energy minimization is a relatively new framework for control, but it has strong ties to stochastic optimal control and encompasses many classical methods, such as proportional-integrative-derivative and linear-quadratic Gaussian control \cite{baioumy2021active,van2021application}.
So far, EFE-based controllers have relied on (partially) known dynamics or neural networks \cite{pio2016active,meera2020free,huebotter2023learning}. 
We propose an EFE minimizing agent based on a polynomial NARX model. NARX models do not require knowledge of system dynamics and require far fewer parameters than neural networks. Furthermore, free energy minimization in autoregressive models has been shown to produce superior parameter estimates at small sample sizes with more stable $k$-step ahead output predictions \cite{fujimoto2016system,kouw2022variational,podusenko2022message}.
%
Our work falls within the scope of information-theoretic model-predictive control \cite{leung2006planning,williams2018information}, but we focus on information gathering specifically for online NARX model identification.

Our contributions include:
\begin{itemize}
    \item The derivation of parameter update rules for a conjugate prior to the Gaussian NARX likelihood. 
    \item The derivation of a location-scale T-distributed posterior predictive distribution for outputs given control inputs.
    \item The derivation of an objective function for an adaptive model-predictive controller that dynamically balances information-seeking and goal-seeking behaviour.
\end{itemize}

\section{Problem statement}
Consider an agent, operating in discrete time, that receives noisy outputs $y_k \in \mathbb{R}$ from a system and sends control inputs $u_k \in \mathbb{R}$ back. It must drive the system to a desired output $y_{*}$ without knowledge of the system's dynamics.

\section{Model specification}
We propose an agent with a probabilistic model that factorizes according to:
\begin{align}
    p( y_{1:N}, u_{1:N}, \theta, \tau) = 
    p(\theta,\tau) \prod_{k=1}^{N} p(y_k \given \theta, \tau, u_k) p(u_k) \, ,
\end{align}
for $N$ time steps. The likelihood is auto-regressive in nature,
\begin{align} \label{eq:likelihood}
    p(y_k \given \theta, \tau, u_k) \triangleq \mathcal{N}(y_k \given \theta^{\intercal} \phi(x_k, u_k), \tau^{-1}) \, ,
\end{align}
with coefficients $\theta \in \mathbb{R}^D$ and measurement noise precision $\tau \in \mathbb{R}^{+}$ \cite{kouw2022variational}. The variable $x_k$ represents the collection of $M_y$ past outputs and $M_u$ past control inputs, $x_k \triangleq (y_{k\-1}, \, \dots y_{k\-M_y}, \, u_{k\-1}, \dots u_{k\-M_u})$. 
%
The function $\phi$ performs a polynomial basis expansion, $\phi : \mathbb{R}^{M} \rightarrow \mathbb{R}^{D}$, where $M = M_y + M_u + 1$ and $D$ is the number of terms in the polynomial.
For example, a second-order polynomial basis without cross-terms with $M_y = M_u = 1$ would produce $\phi(y_{k\-1}, u_{k\-1}, u_k) = [1 \ y_{k\-1} \ u_{k\-1} \ u_k \ y_{k\-1}^2 \ u_{k\-1}^2 \ u_k^{2}]$. 
%

The prior distribution on the parameters is a multivariate Gaussian - univariate Gamma distribution \cite[ID: D5]{Soch}:
\begin{align} \label{eq:priors}
    p(\theta, \tau) &\triangleq \mathcal{NG}\big(\theta, \tau \given \mu_0, \Lambda_0, \alpha_0, \beta_0 \big) \\
    &= \mathcal{N}(\theta \given \mu_0, (\tau \Lambda_0)^{-1} \big) \, \mathcal{G}\big(\tau \given \alpha_0, \beta_0 \big) \, .
\end{align}
This choice of parameterization allows for inferring an exact posterior distribution (Lemma III.1) and an exact posterior predictive distribution (Lemma III.2). 
%

The prior distributions over control inputs are assumed to be independent over time:
\begin{align}
    p(u_k) \triangleq \cN(u_k \given 0, \eta^{-1}) \label{eq:control-prior} \, ,
\end{align}
with precision parameter $\eta$. This choice has a regularizing effect on the inferred controls (Sec.~\ref{sec:fe-controls}). 

\section{Inference} 

\subsection{Parameter estimation} \label{sec:fe-params}
First, we note that, at time $k$, the control $u_k$ has been executed and is known to the agent. Henceforth, we shall use $\hat{u}_k$ and $\hat{y}_k$ to differentiate observed variables from unobserved ones, i.e., $u_k$ and $y_k$.

The parameter posterior distribution is obtained by Bayesian filtering \cite{sarkka2013bayesian}:
\begin{align} \label{eq:bayesian-filtering}
    \underbrace{p\big(\theta, \tau \given \mathcal{D}_{k} \big)}_{\text{posterior}} \! = \! \frac{\overbrace{p\big(\hat{y}_k \given \theta, \tau, \hat{u}_k \big)}^{\text{likelihood}}}{\underbrace{p\big(\hat{y}_{k} \given \hat{u}_{k}, \mathcal{D}_{k\-1}\big)}_{\text{evidence}}} \underbrace{p\big(\theta, \tau \given \mathcal{D}_{k\-1}\big)}_{\text{prior}} . 
\end{align}
where $\mathcal{D}_k = \{\hat{y}_i, \hat{u}_i\}_{i=1}^{k}$ is the data up to time $k$. The evidence (a.k.a. marginal likelihood) is
\begin{align} \label{eq:evidence-filtering}
    p\big(\hat{y}_{k} | \hat{u}_{k}, \mathcal{D}_{k\-1}\big) \! = \! \int p\big(\hat{y}_{k} | \theta, \tau, \hat{u}_k \big) p\big(\theta,\tau | \mathcal{D}_{k\-1}\big) \, \mathrm{d}(\theta,\tau) .
\end{align}
This integral is typically intractable. In earlier work, we obtained an approximate posterior distribution through mean-field variational inference \cite{kouw2022variational}. 
In this model we obtain an exact posterior distribution using the multivariate Gaussian - univariate Gamma prior distribution specified in Eq.~\ref{eq:priors}.
\begin{lemma} \label{eq:lemma1}
    The Gaussian distributed NARX likelihood in Eq.~\ref{eq:likelihood} combined with a multivariate Gaussian - univariate Gamma prior distribution over AR coefficients $\theta$ and measurement precision $\tau$, 
    \begin{align}
        p(\theta, \tau \given \mathcal{D}_{k\-1}) = \mathcal{NG}(\theta, \tau \given \mu_{k\-1}, \Lambda_{k\-1}, \alpha_{k\-1}, \beta_{k\-1}) \, ,
    \end{align}
    yields a multivariate Gaussian - univariate Gamma posterior distribution
    \begin{align} \label{eq:truepost}
        p(\theta, \tau \given \mathcal{D}_k) = \mathcal{NG}(\theta, \tau \given \mu_k, \Lambda_k, \alpha_k, \beta_k) \, .
    \end{align}
    The parameters of the posterior distribution are
    \begin{align} \label{eq:truepost_params}
        &\mu_k = \big(\phi_k \phi_k^{\intercal} \! + \! \Lambda_{k\-1} \big)^{-1}\big(\phi_k \hat{y}_k \! + \! \Lambda_{k\-1} \mu_{k\-1} \big) , \\ 
        &\Lambda_k = \phi_k \phi_k^{\intercal}  +  \Lambda_{k\-1} , \ \qquad \quad
        \alpha_k = \alpha_{k\-1} + \frac{1}{2} , \nonumber \\
        &\beta_k = \beta_{k\-1} +  \frac{1}{2}\big( \hat{y}_k^2  -  \mu_k^{\intercal} \Lambda_k \mu_k + \mu_{k\-1}^{\intercal}\Lambda_{k\-1} \mu_{k\-1} \big) , \nonumber 
    \end{align}
    where $\phi_k \triangleq \phi(\hat{x}_k, \hat{u}_k)$.
\end{lemma}
The proof is in \nameref{sec:appendix-A}. 
%
The marginal posterior distributions are Gamma distributed and multivariate location-scale T-distributed \cite[ID: P36]{Soch}:
\begin{align}
    p(\tau \given \mathcal{D}_k)  &= \! \int \! p(\theta, \tau | \mathcal{D}_k) \mathrm{d}\theta = \mathcal{G}(\tau \given \alpha_k, \beta_k) \label{eq:margpost-tau} \, , \\
    p(\theta \given \mathcal{D}_k)  &= \! \int \! p(\theta, \tau | \mathcal{D}_k) \mathrm{d}\tau = \mathcal{T}_{2\alpha_k}\big(\theta \given \mu_k, \frac{\beta_k}{\alpha_k} \Lambda_k^{\-1}\big) . \label{eq:margpost-theta}
\end{align}
The subscript under $\mathcal{T}$ refers to its degrees of freedom.

\subsection{Control estimation} \label{sec:fe-controls}
%
In order to effectively drive the system to the goal output, the agent must make accurate predictions for future outputs. We express the probability of the output and parameters at time $t = k+1$ given a future control input as
\begin{align} \label{eq:p-future}
    p(y_t, \theta, \tau \given u_t) = p(y_t \given \theta, \tau, u_t) \,  p(\theta, \tau \given \mathcal{D}_k) \, .
\end{align}
The dependence on $\mathcal{D}_k$ on the left-hand side is left implicit for the sake of brevity.
Note that this probability distribution does not yet include a goal output. Without it, the agent will drive the system towards \emph{any} output that leads to minimal prediction error. To incorporate the goal output, we first decompose the joint distribution above to:
\begin{align} \label{eq:decomp-futurejoint}
    p(y_t, \theta, \tau \given u_t) = p(\theta, \tau \given y_t, u_t) \, p(y_t) \, .
\end{align}
Then, we constrain the marginal prior distribution over future output, $p(y_t)$, to a specific functional form with chosen parameters:
\begin{align} 
    p(y_t \given y_{*}) \triangleq \cN(y_t \given m_{*}, v_{*}) \, . \label{eq:goalp}
\end{align}
This constrained distribution on the future output is known as a \emph{goal prior} distribution \cite{friston2015active}.

Using Eqs.~\ref{eq:decomp-futurejoint}, \ref{eq:goalp} and Bayes' rule, the posterior distribution for the controls is:
\begin{align}
    p(u_t \given y_t, \theta, \tau) = \frac{p(\theta, \tau \given y_t, u_t) p(y_t \given y_{*}) p(u_t)}{\int p(\theta, \tau \given y_t, u_t) p(y_t \given y_{*}) p(u_t) \mathrm{d}u_t } \, .
\end{align}
The integral in the denominator is challenging to solve due to the basis function applied to $u_t$ inside the likelihood. Instead, we introduce a variational distribution $q(u_t)$ to approximate the posterior.
The approximation error is characterized with an expected free energy functional \cite{friston2015active,van2022active},
\begin{align} \label{eq:EFE0}
\mathcal{F}_t[q]  \triangleq   \mathbb{E}_{q(y_t, \theta, \tau, u_t)} \Big[\! \ln \frac{p(\theta, \tau \given \mathcal{D}_k)q(u_t)}{p(\theta, \tau | y_t, u_t) p(y_t | y_{*}) p(u_t)} \Big] ,
\end{align}
where the expectation is over
\begin{align} \label{eq:q-factorization}
    q(y_t, \theta, \tau, u_t) \triangleq p(y_t, \theta, \tau \given u_t) q(u_t) \, .
\end{align}

Inferring the optimal control at time $t$ refers to minimizing the free energy functional with respect to the variational distribution $q(u_t)$:
\begin{align}
    q^{*}(u_t) = \underset{q \, \in \, Q}{\arg \min} \ \mathcal{F}_t[q] \, .
\end{align}
where $Q$ represents the space of candidate distributions.
We can re-arrange the free energy functional to simplify the variational minimization problem:
\begin{align}
&\mathbb{E}_{q(y_t, u_t, \theta, \tau)} \Big[ \ln \frac{p(\theta, \tau \given \mathcal{D}_k) \, q(u_t)}{p(\theta, \tau \given y_t, u_t) p(y_t \given y_{*}) p(u_t)} \Big] = \\
& \! \mathbb{E}_{q(u_t)} \Big[ \mathbb{E}_{p(y_t, \theta, \tau | u_t)} \big[ \ln \frac{p(\theta, \tau \given \mathcal{D}_k)}{p(\theta, \tau | y_t, u_t) p(y_t | y_{*})} \big] \! + \! \ln \frac{q(u_t)}{p(u_t)} \Big] . \nonumber
\end{align}
Now we define the expected free energy \emph{function}:
\begin{align} \label{eq:EFE1}
    \cJ(u_t) \triangleq \mathbb{E}_{p(y_t, \theta, \tau \given u_t)} \Big[ \ln \frac{p(\theta, \tau \given \mathcal{D}_k)}{p(\theta, \tau \given y_t, u_t) p(y_t \given y_{*})} \Big] \, .
\end{align}
Using $\cJ(u_t) = \ln (1/ \exp(-\cJ(u_t)))$, the expected free energy functional can be expressed as a Kullback-Leibler divergence
\begin{align} \label{eq:EFE2}
    \cF_t[q] = \mathbb{E}_{q(u_t)} \Big[ \ln \frac{q(u_t)}{\exp\big(-\cJ(u_t) \big) p(u_t)} \Big] \, ,
\end{align}
%
%
which is minimal when \cite{mackay2003information}:
\begin{align} \label{eq:optqu}
    q^{*}(u_t) =  \exp\big(- \cJ(u_t) \big) p(u_t)\, .
\end{align}
Thus, we have an optimal approximate posterior distribution over controls. 

To proceed, we must solve the expectation in \eqref{eq:EFE1}. 
We have access to the parametric forms of all distributions involved except for the distribution over parameters given the future output and control. It can be tied to known distributions through Bayes' rule:
\begin{align} 
    p(\theta, \tau &\given y_t, u_t) =  \frac{p(y_t \given \theta, \tau, u_t)\ p(\theta, \tau \given \mathcal{D}_k)}{\int p(y_t \given \theta, \tau, u_t) p(\theta,\tau \given \mathcal{D}_k) \mathrm{d}(\theta,\tau)} . \label{eq:futureparampost-bayes}
\end{align}
The distribution that results from the marginalization in the denominator is the posterior predictive distribution $p(y_t \given u_t)$ and can be derived analytically within our model.
\begin{lemma} \label{eq:lemma2}
    The marginalization of the NARX likelihood (Eq.~\ref{eq:likelihood}) over the parameter posterior distribution (Eq.~\ref{eq:truepost}) yields a location-scale T-distribution:
    \begin{align} 
    &p(y_t \given u_t) = \int p(y_t \given \theta, \tau, u_t) \ p(\theta, \tau \given \mathcal{D}_{k}) \, \mathrm{d}(\theta, \tau) \\
    &= \! \int \! \cN(y_t | \theta^{\intercal} \phi_t, \tau^{-1}) \mathcal{NG}( \theta, \tau | \mu_k, \Lambda_k, \alpha_k, \beta_k) \mathrm{d}(\theta, \tau) \label{eq:postpred-GNG} \\
    &= \ \mathcal{T}_{\nu_t} \Big(y_t \given m_t,  \ s^2_t \Big) \, , \label{eq:q-postpred}
\end{align}
where $\phi_t \triangleq \phi(\hat{x}_t, u_t)$ and
\begin{align} \label{eq:shorthand_post}
    \nu_t \triangleq 2\alpha_k \, , \ m_t \triangleq \mu_k^{\intercal} \phi_t \, , \ s^2_t \triangleq \frac{\beta_k}{\alpha_k}\big(\phi_t^{\intercal} \Lambda_k^{-1} \phi_t + 1\big) . 
\end{align}
\end{lemma}
The proof is in \nameref{sec:appendix-B}. 

If we replace $p(\theta, \tau \given y_t, u_t)$ in Eq.~\ref{eq:EFE1} with the right-hand side of Eq.~\ref{eq:futureparampost-bayes} and use Eq.~\ref{eq:p-future}, then the EFE function can be split into two components:
\begin{align}
\cJ(u_t) 
    &= \mathbb{E}_{p(y_t, \theta, \tau \given u_t)} \Big[\ln \frac{1}{p(y_t \given y_{*})} \Big] \label{eq:EFE3}  \\
    &\quad + \mathbb{E}_{p(y_t, \theta, \tau \given u_t)} \Big[\ln \frac{p(\theta, \tau \given \mathcal{D}_k) p(y_t \given u_t)}{p(y_t \given \theta, \tau, u_t) p(\theta, \tau \given \mathcal{D}_k)} \Big] \nonumber \\
    &= \mathbb{E}_{p(y_t \given u_t)} \Big[-\ln p(y_t \given y_{*}) \Big] \label{eq:EFE4} \\
    &\quad - \mathbb{E}_{p(y_t, \theta, \tau \given u_t)} \Big[\ln \frac{p(y_t, \theta, \tau \given u_t) }{p(\theta, \tau \given \mathcal{D}_k) p(y_t \given u_t)} \Big] \nonumber \, .
\end{align}
One may recognize the first term as a cross-entropy, describing the dissimilarity between the predictive distribution and the goal prior distribution \cite{mackay2003information}. The second term is the mutual information between the parameter posterior and the posterior predictive distribution. It describes how much information is gained on the parameters upon measuring a system output \cite{mackay2003information}.
Solving the expectations yields a compact objective:
\begin{theorem}
The expected free energy function in Eq.~\ref{eq:EFE4} evaluates to:
\begin{align} \label{eq:EFE-final}
    \cJ(u_t) &= \frac{1}{2 v_{*}} \Big(\big(\mu_k^{\intercal} \phi_t - m_{*} \big)^2 \! + \frac{\beta_k}{\alpha_k}\big(\phi_t^{\intercal} \Lambda_k^{-1} \phi_t \! + \! 1\big) \frac{2\alpha_k}{2\alpha_k \! - \! 2} \Big) \nonumber \\
    &\qquad - \frac{1}{2} \ln \Big(\phi_t^{\intercal}\Lambda_k^{-1}\phi_t + 1\Big) + \text{constants} \, . 
\end{align}
\end{theorem}
The proof is found in \nameref{sec:appendix-C}. 

We consider MAP estimation of the control policy because, although it is informative to quantify uncertainty over controls, we can ultimately only execute one action. The MAP estimate can be expressed as a minimization over a negative logarithmic transformation of $q^{*}(u_t)$ (Eq.~\ref{eq:optqu}):
\begin{align} \label{eq:neglog}
    \hat{u}_t = \underset{u_t \, \in \, \mathcal{U}}{\arg \max} \, q^{*}(u_t) =  \underset{u_t \, \in \, \mathcal{U}}{\arg \min} \ \cJ(u_t) - \ln p(u_t) \, .
\end{align}
The $\mathcal{U}$ refers to the space of affordable controls and serves to incorporate practical constraints such as torque limits.
%

So far, we have only considered a $1$-step ahead prediction. Generalizing to $t > k+1$ is, in principle, straightforward. Due to the independence assumptions on the prior $p(u_t)$ and the variational control posteriors $q(u_t)$, the joint variational control posterior distribution factorizes over time:
\begin{align}
    q^{*}(u_t, \dots, u_{t+T}) = \prod_{t=1}^T \, p(u_t) \, \exp\big(-\cJ(u_t) \big) \, .
\end{align}
If we apply the same negative logarithmic transformation as in Eq.~\ref{eq:neglog} to MAP estimation for the joint variational control posterior, then the final control optimization problem becomes:
\begin{align} \label{eq:u_EFE}
    \hat{u}^{\text{EFE}} =&\ \underset{u \, \in \, \mathcal{U}^{T}}{\arg \min} \ \sum_{t=1}^T \frac{1}{2 v_{*}} \big(\mu_k^{\intercal} \phi(\hat{x}_t, u_t) \! - \! m_{*} \big)^2 \\
    &  + \! \frac{\beta_k}{v_{*}(2\alpha_k-2)} \Big(\big(\phi(\hat{x}_t, u_t)^{\intercal} \Lambda_k^{-1} \phi(\hat{x}_t, u_t) \! + \! 1\big) \Big) \nonumber \\
    &  - \frac{1}{2} \ln \Big(\phi(\hat{x}_t, u_t)^{\intercal}\Lambda_k^{-1}\phi(\hat{x}_t, u_t) \! + \! 1\Big) + \eta u_t^2 \, . \nonumber 
\end{align}
where $u = (u_t, \dots, u_T)$ and $\mathcal{U}^T$ is the Cartesian product of $T$ spaces over affordable controls.

However, this generalization to $t>k+1$ assumes that all elements of $x_t$ have been observed. The predicted $y_t$ is typically incorporated into $x_{t+1}$ as if it were an observation \cite{billings2013nonlinear,khandelwal2018simulation}. However, from a Bayesian perspective, $y_t$ is not a number but a random variable. This causes a problem because the delay vector $x_t$ would also become a random variable for $t > k+1$, and would interfere with the conditioning and marginalization operations described earlier. We avoid this problem by collapsing the posterior predictive distribution for $y_t$ (Eq.~\ref{eq:q-postpred}) into its most probable value $\mu_k^{\intercal}\phi(\hat{x}_t, \hat{u}_t)$, and incorporating that number into $x_{t+1}$. This is a suboptimal solution because the predictive variance will not accumulate over the horizon. 

The optimization problem in Eq.~\ref{eq:u_EFE} can be solved by iterative methods with automatic differentiation for obtaining the gradient. The box constraints over affordable controls $\mathcal{U}$ can be incorporated by utilizing an interior-point method.




\section{Experiments} \label{sec:experiments}

\subsection{Seeking informative observations}
The more informative a system input-output observation, the faster the agent is able to accurately estimate parameters and make accurate predictions. 
We shall compare an agent based on the EFE objective in Eq.\ref{eq:u_EFE} (referred to as EFE) with an agent that uses the typical quadratic cost with regularization objective (referred to as QCR) \cite{grune2017nonlinear}:
\begin{align} \label{eq:u_QCR}
    \hat{u}^{\text{QCR}} = \underset{ u \, \in \, \mathcal{U}^T}{\arg \min} \sum_{t=1}^T \, \big(\mu_k^{\intercal}\phi(\hat{x}_t, u_t) - m_{*} \big)^2 + \eta u_t^2 \, .
\end{align}
QCR can be described as purely goal-seeking, because it only seeks to minimize the difference between the predictions and the goal. When comparing EFE with QCR we observe the effect of adding parameter uncertainty and predictive variance to the objective function.

\begin{figure}[htb]
    \includegraphics[width=.48\textwidth]{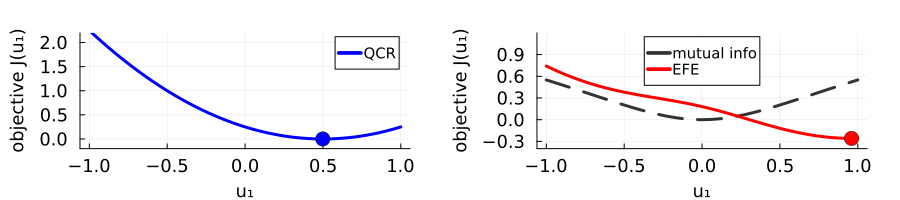} \\
    \includegraphics[width=.48\textwidth]{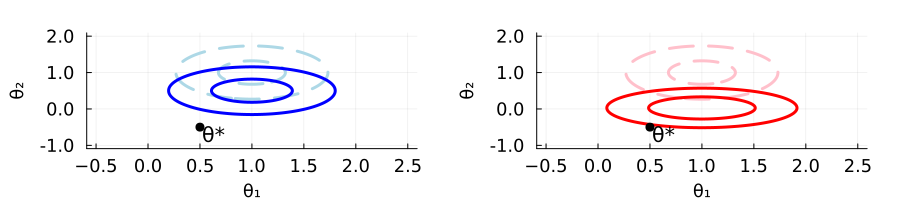} \\
    \includegraphics[width=.48\textwidth]{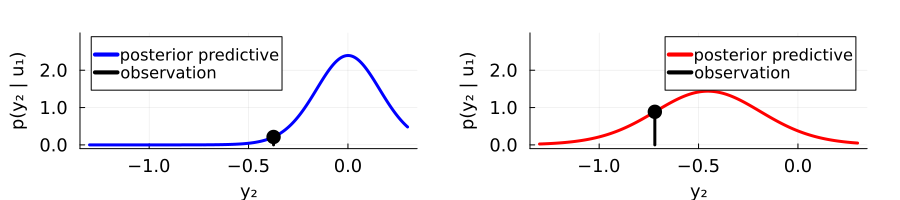} \vspace{-3pt}
    \caption{Example of information-seeking. (Top row) Control objective functions for quadratic cost (QCR, left) vs proposed cost (EFE, right). Black dashed line displays mutual information, which causes $\hat{u}_1^{\text{EFE}} > \hat{u}_2^{\text{QCR}}$. (Middle row) Marginal prior (dashed) and posterior (solid) distributions for coefficients $\theta$ obtained after executing $\hat{u}_1^{\text{QCR}}$ (left) and $\hat{u}_1^{\text{EFE}}$ (right), and observing the resulting $\hat{y}_1$'s. Note that the posteriors under EFE controls are closer to the system AR coefficients $\theta^{*}$. (Bottom row) Posterior predictive distributions for $y_2$ under QCR (left) and EFE (right) controls. Note that $\hat{y}_2$'s probability is larger under EFE, i.e., the prediction error is smaller.}
    \label{fig:sbs-objectives}
    \vspace{-2pt}
\end{figure}

Consider a system that evolves according to $y_k = \theta^{*}_1 y_{k\-1} + \theta^{*}_2 u_k$ where $\theta^{*}_1 = 0.5$ and $\theta^{*}_2 = -0.5$.
The task is to drive this system to $m_{*} = 0.5$ (for EFE, $v_{*}=1$) under $\mathcal{U} = [-1 \, 1]$ for $T$ = $1$. We used Optim.jl to minimize the objective functions \cite{mogensen2018optim}. The models' prior parameters are set to shape $\alpha_0$ = $10$, rate $\beta_0$ = $1$, mean $\mu_0$ = $[1 \ 1]$, precision $\Lambda_0$ = $\frac{1}{2}I$ and control precision $\eta=0$.
The top row in Figure \ref{fig:sbs-objectives} compares the two objective functions (QCR left, EFE right) for the first action $u_1$, with $\hat{u}^{\text{QCR}}_1 = 0.5$ and $\hat{u}^{\text{EFE}}_1 = 0.96$. This difference is caused by the mutual information term (dashed gray), which indicates that $u_t$'s further from $0$ are more informative (note Eq.~\ref{eq:u_EFE} uses \emph{negative} mutual information, meaning the objective is smaller for more informative controls). After executing the respective $\hat{u}_1$'s, the systems return $\hat{y}_1$'s and the parameter distributions are updated. The middle row in Figure \ref{fig:sbs-objectives} shows contour levels of the marginal prior and posterior parameter distributions (see Eq.~\ref{eq:margpost-theta}) over the $[\theta_1 \, \theta_2]$ plane. Note that the posterior under the EFE objective has moved closer to the system coefficients $\theta^{*}$ (i.e., $p(\theta^{*} \given \mathcal{D}_1)$ is higher). The bottom row in Figure \ref{fig:sbs-objectives} show the posterior predictive distributions for $y_2$ and indicates that EFE's predictions are closer to the systems' $\hat{y}_2$. 

\begin{figure}[htb]
    \includegraphics[width=.48\textwidth]{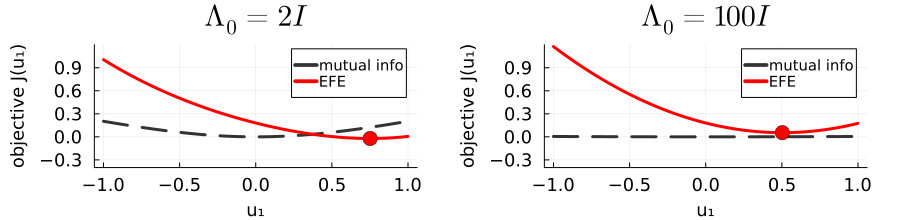}
    \vspace{-5pt}
    \caption{Examples of how parameter certainty affects the mutual information term and the resulting EFE control objective. For $\Lambda_0 = 2I$ (left; medium certainty) the minimizer shrinks to $u_1 = 0.75$ (compared to $\Lambda_0 = \frac{1}{2}I$, Figure \ref{fig:sbs-objectives} top right). For $\Lambda_0 = 100I$ (right; high certainty), the minimizer shrinks all the way to $u_1 = 0.5$ as in the QCR objective (Figure \ref{fig:sbs-objectives} top left).}
    \label{fig:sbs-lowuncertainty}
    \vspace{-15pt}
\end{figure}

The information-seeking quality depends on parameter certainty, as can be seen by replicating the previous example with different values of $\Lambda_0$. Figure \ref{fig:sbs-lowuncertainty} (left) shows that the mutual information term is flatter for $\Lambda_0 = 2I$. It has less of an effect on the overall EFE objective (compare with $\Lambda_0 = \frac{1}{2}I$ shown in Figure \ref{fig:sbs-objectives} top right) and the optimal control input shrinks to $u_1 = 0.75$. For $\Lambda_0 = 100 I$ (see Figure \ref{fig:sbs-lowuncertainty} right), the mutual information term is so flat that the EFE objective resembles the QCR objective (see Figure \ref{fig:sbs-objectives} top left), and is also minimal at $u_1 = 0.5$.

\subsection{Control under unknown dynamics}
In this experiment, the controllers must swing up a damped pendulum and maintain an upright position\footnote{Code at \url{https://github.com/biaslab/LCSS2024-NARXEFE}}. 
The system parameters are mass $\mathrm{m} = 1.0$, length $\ell = 0.5$, friction $\mathrm{c} = 0.01$ and time-step $\Delta t = 0.1$ . The angle $\vartheta_t$ is observed under zero-mean Gaussian noise with a standard deviation of $0.001$.
The agent's NARX model consists of a second-order polynomial basis without cross-terms and with delays $M_y = 2$, $M_u = 2$. The prior distribution's parameters are shape $\alpha_0 = 10$, rate $\beta_0 = 0.1$, mean $\mu_0 = 10^{-8} \cdot [1 \dots 1]$, coefficient precision $\Lambda = \frac{1}{2}I$ and control prior precision $\eta = 0.001$. These parameters would be regarded as weakly informative for noise precision $\tau$ and non-informative for coefficients $\theta$. The goal prior has mean $m_{*} = \pi$ (i.e., upward) and variance $v_{*} = 0.5$.
\begin{figure}[thb]
    \includegraphics[width=0.48\textwidth]{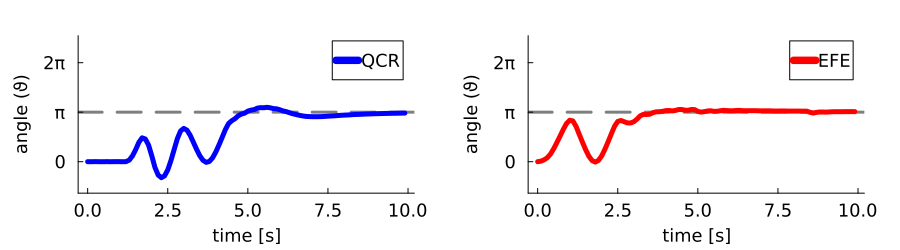}
    \includegraphics[width=0.48\textwidth]{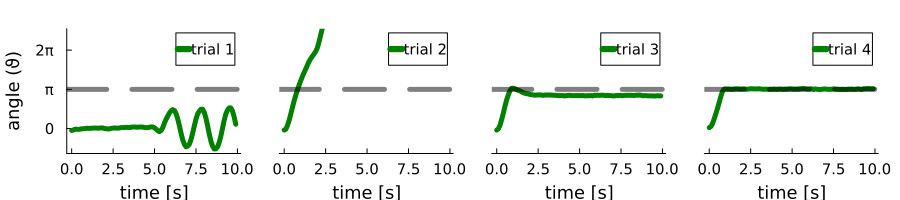}
    \vspace{-10pt}
    \caption{Pendulum angle over time during swing-up task under unknown dynamics. (Top left) QCR-based controller, (top right) EFE-based controller. (Bottom) Trials for PILCO-based controller.}
    \label{fig:exp-pendulumswings}
    \vspace{-5pt}
\end{figure}
The top row of Figure \ref{fig:exp-pendulumswings} visualizes the paths the two controllers took. The QCR-based controller does not initially move much as it cannot predict the effect of the control inputs under the non-informative prior for the coefficients. When it starts, it swings a few times before it reaches the goal and settles. The EFE-based controller moves away from zero earlier (i.e., it is seeking system input-output observations that produce more accurate output predictions) and swings up the pendulum sooner.

For the sake of comparison, we also report the performance of PILCO \cite{deisenroth2011pilco}. It is a Gaussian Process-based identification method and although it is one of the most data-efficient reinforcement learning methods, it still requires several trials before it finds a successful policy. The bottom row of Figure \ref{fig:exp-pendulumswings} shows its progress over the first 4 trials.


\section{Discussion} \label{sec:discussion}


Close inspection of Eq.~\ref{eq:EFE-final} reveals that the balance between the mutual information, cross-entropy and prior distribution is also affected by the goal prior variance $v_{*}$. It may be possible to optimize this dynamically.
Furthermore, note that the optimization problem in Eq.~\ref{eq:u_EFE} is quadratic in $\phi_t$. This means that for all nonlinear polynomials (i.e., degree larger than 1), the optimization becomes non-convex in $u_t$.

%

\section{Conclusion}
We have proposed an adaptive NARX model-predictive controller that seeks informative observations for online system identification when its parameter uncertainty is high, and seeks control inputs that drive the system to a goal when its parameter uncertainty is low. This is achieved through a control objective with information-theoretic quantities, derived from the free energy minimization framework. 

\section*{Appendix A}\label{sec:appendix-A}
\begin{proof}
    The posterior is proportional to the likelihood times the prior distribution:
    \begin{align}
        &p(\theta, \tau \given \mathcal{D}_{k}) \nonumber \\
        &\propto \cN\big(\hat{y}_k \given \theta^{\intercal}\phi_k, \tau^{-1} \big) \mathcal{NG}\big(\theta, \tau | \mu_{k\-1}, \Lambda_{k\-1}, \alpha_{k\-1}, \beta_{k\-1}) \\
        &\propto \tau^{1/2} \exp\big(-\frac{\tau}{2}(\hat{y}_k - \theta^{\intercal}\phi_k)^2\big) \label{eq:truepost1} \\
        &\ \ \tau^{D/2+\alpha_{k\-1} -1} \! \exp\big( \! \! - \! \frac{\tau}{2} ((\theta \! - \! \mu_{k\-1})^{\intercal} \Lambda_{k\-1} (\theta \! - \! \mu_{k\-1}) \! + \! 2\beta_{k\-1})\big) \nonumber \\
        &= \tau^{\alpha_{k\-1} + (D+1)/2 -1} \exp \Big(  -  \frac{\tau}{2} \big( (\hat{y}_k  -  \theta^{\intercal}\phi_k)^2 \label{eq:truepost2} \\
        & \qquad \qquad \qquad \quad +  (\theta  -  \mu_{k\-1})^{\intercal} \Lambda_{k\-1} (\theta  -  \mu_{k\-1}) +  2\beta_{k\-1} \big)  \Big) . \nonumber
    \end{align}
    Expanding the squares within the exponential and collecting terms involving $\theta$ yields:
    \begin{align}
        (\hat{y}_k & -  \theta^{\intercal}\phi_k)^2 + (\theta - \mu_{k\-1})^{\intercal} \Lambda_{k\-1} (\theta  -  \mu_{k\-1})   \nonumber \\
        &= \theta^{\intercal} \big(\Lambda_{k\-1}  +  \phi_k \phi_k^{\intercal} \big) \theta - \big(\mu_{k\-1}^{\intercal} \Lambda_{k\-1}  +  \hat{y}_k \phi_k^{\intercal} \big) \theta \\
        & \quad - \theta^{\intercal} \big(\Lambda_{k\-1} \mu_{k\-1} + \phi_k \hat{y}_k \big) + \mu_{k\-1}^{\intercal}\Lambda_{k\-1} \mu_{k\-1} + \hat{y}_k^2  \nonumber \, .
    \end{align}
    Then, using $\Lambda_k \triangleq \Lambda_{k\-1} + \phi_k \phi_k^{\intercal}$ and $\xi_k \triangleq \Lambda_{k\-1} \mu_{k\-1} + \phi_k \hat{y}_k$, one may isolate a quadratic form:
    \begin{align}
        \theta^{\intercal} \Lambda_k \theta &- \xi_k^{\intercal} \theta  - \theta^{\intercal} \xi_k   \label{eq:truepost3} \\
        &= \big(\theta \! - \! \Lambda_k^{-1} \xi_k \big)^{\intercal} \Lambda_k \big(\theta \! - \! \Lambda_k^{-1} \xi_k \big) - \xi_k^{\intercal} \Lambda_k^{-1} \xi_k  \nonumber \, .
    \end{align}
Plugging the above back into Eq.~\ref{eq:truepost2}, reveals a multivariate Gaussian - univariate Gamma distribution:
\begin{align}
    p(\theta, \tau \given \mathcal{D}_k) &\propto \tau^{\alpha_{k\-1} + (D+1)/2 -1}  \\
    &\quad \exp\Big(\! - \! \frac{\tau}{2} \big( (\theta \! - \! \Lambda_k^{-1} \xi_k )^{\intercal} \Lambda_k \big(\theta \! - \! \Lambda_k^{-1} \xi_k \big)  \nonumber \\
    &\quad - \xi_k^{\intercal} \Lambda_k^{-1} \xi_k + \mu_{k\-1}^{\intercal}\Lambda_{k\-1} \mu_{k\-1} + \hat{y}_k^2  + 2\beta_{k\-1} \big) \Big) \nonumber \\
    &= \mathcal{NG}(\theta, \tau \given \mu_k, \Lambda_k, \alpha_k, \beta_k) \, ,
\end{align}
with $\Lambda_k$ defined as above \eqref{eq:truepost3}, $\alpha_k \triangleq \alpha_{k\-1} + 1/2$ and
\begin{align}
    &\mu_k \triangleq \Lambda_k^{-1}\xi_k = (\Lambda_{k\-1} + \phi_k \phi_k^{\intercal})^{-1}(\Lambda_{k\-1}\mu_{k\-1} + \phi_k \hat{y}_k) \, , \nonumber \\ 
    & \beta_k \triangleq \beta_{k\-1} +  \frac{1}{2}\big( \hat{y}_k^2  -  \xi_k^{\intercal} \Lambda_k^{-1} \xi_k + \mu_{k\-1}^{\intercal}\Lambda_{k\-1} \mu_{k\-1} \big) . 
\end{align}
Note that $\xi_k = \Lambda_k \mu_k$, so $\xi_k^{\intercal} \Lambda_k^{-1} \xi_k = \mu_k^{\intercal} \Lambda_k \mu_k$.
\end{proof}

\section*{Appendix B} \label{sec:appendix-B}
\begin{proof}
    We start by combining the Gaussian likelihood with the conditional Gaussian distribution for $\theta$:
    \begin{align} 
        &\cN \big( y_t \given \theta^{\intercal} \phi_t, \tau^{-1} \big) \mathcal{N}\big( \theta \given \mu_k, (\tau \Lambda_k)^{-1}  \big) = \label{eq:postpred_pytau} \\
        &\  \cN\Big( \begin{bmatrix} \theta \\ y_t \end{bmatrix} \big| \begin{bmatrix} \mu_k \\ \mu_k^{\intercal}\phi_t \end{bmatrix} \! , \! \begin{bmatrix} (\tau\Lambda_k)^{-1} & (\tau\Lambda_k)^{-1}\phi_t \\ \phi_t^{\intercal} (\tau\Lambda_k)^{-1} & \phi_t^{\intercal}(\tau\Lambda_k)^{-1} \phi_t + \tau^{-1}  \end{bmatrix} \Big) . \nonumber
    \end{align}
    Marginalizing \eqref{eq:postpred_pytau} over $\theta$ yields \cite[ID: P35]{Soch}:
    \begin{align}
        p(y_t \given \tau, u_t) = \cN\big(y_t \given \mu_k^{\intercal}\phi_t, \tau^{-1}(\phi_t^{\intercal}\Lambda_k^{-1} \phi_t + 1) \big) \, .
    \end{align}
    Let $m_t \triangleq \mu_k^{\intercal}\phi_t$ and $\lambda_t \triangleq (\phi_t^{\intercal} \Lambda_k^{-1} \phi_t + 1)^{-1}$. Then
    \begin{align}
        &p(y_t \given u_t) = \int p(y_t \given \tau, u_t) p(\tau) \mathrm{d}\tau \\
        &= \! \int \! \cN\big(y_t \given m_t, (\tau \lambda_t)^{-1} \big) \mathcal{G}(\tau \given \alpha_k, \beta_k) 
        \mathrm{d}\tau \\
        &= \! \int \big(\frac{\lambda_t}{2\pi}\big)^{1/2} \tau^{1/2} \exp\big(\! - \! \frac{\tau \lambda_t}{2}(y_t - m_t)^2\big)  \\
        &\qquad \qquad \qquad \qquad \quad  \frac{\beta_k^{\alpha_k}}{\Gamma(\alpha_k)} \tau^{\alpha_k-1} \exp(-\tau \beta_k) \ \mathrm{d}\tau  \nonumber \\
        &= \! \big(\frac{\lambda_t}{2\pi}\big)^{1/2} \frac{\beta_k^{\alpha_k}}{\Gamma(\alpha_k)} \label{eq:postpred_int1} \\
        &\qquad \int \tau^{\alpha_k + 1/2 - 1}
        \exp\Big(\! - \! \tau \big(\beta_k + \frac{ \lambda_t}{2}(y_t - m_t)^2\big) \Big) \ \mathrm{d}\tau  \nonumber \\
        &= \! \big(\frac{\lambda_t}{2\pi}\big)^{1/2} \frac{\beta_k^{\alpha_k}}{\Gamma(\alpha_k)} \frac{\Gamma(\alpha_k + \frac{1}{2})}{\Big(\beta_k + \frac{ 1}{2}\lambda_t(y_t - m_t)^2 \Big)^{\alpha_k + 1/2}} \label{eq:postpred_int2} \\
        &= \! \big(\frac{\lambda_t}{2\pi}\big)^{1/2}\frac{\Gamma(\alpha_k \! + \! \frac{1}{2})}{\Gamma(\alpha_k)} \Big(\frac{\beta_k}{\beta_k \! + \! \frac{ 1}{2}\lambda_t(y_t \! - \! m_t)^2} \Big)^{\alpha_k + 1/2} \beta_k^{-1/2} \\
        &= \! \big(\frac{\alpha_k \lambda_t}{\alpha_k \beta_k 2\pi}\big)^{1/2}\frac{\Gamma(\frac{2\alpha_k + 1}{2})}{\Gamma(\frac{2\alpha_k}{2})} \! \big(1 \! + \! \! \frac{ \alpha_k \lambda_t (y_t \! - \! m_t)^2}{2 \alpha_k \beta_k} \! \big)^{-(2\alpha_k \! + \! 1)/2}
    \end{align}
    where the integrand in \eqref{eq:postpred_int1} is an unnormalized Gamma distribution and evaluates to its normalization constant in \eqref{eq:postpred_int2}.
    Defining $\nu_t \triangleq 2\alpha_k$ and $s^2_t \triangleq (\frac{\alpha_k}{\beta_k} \lambda_t)^{-1}$, reveals a location-scale T-distribution:
    \begin{align}
        p(y_t | u_t) \! = \! \frac{\Gamma(\frac{\nu_t + 1}{2})}{\sqrt{\pi \nu_t s^2_t }\, \Gamma(\frac{\nu_t}{2})} \big(1 \! + \!  \frac{(y_t  \! - \! m_t)^2}{\nu_t s^2_t} \big)^{-(\nu_t + 1)/2} .
    \end{align}
\end{proof}

\section*{Appendix C} \label{sec:appendix-C}
\begin{proof}
The cross-entropy in \ref{eq:EFE4} is:
\begin{align}
    &\mathbb{E}_{\mathcal{T}_{\nu_t}(y_t \given m_t, s_t^2)} \Big[-\ln \cN(y_t \given m_{*}, v_{*}) \Big]  \nonumber \\
    &= \frac{1}{2}\ln (2\pi v_{*}) + \frac{1}{2v_{*}} \Big( (m_t - m_{*})^2 + s^2_t\frac{\nu_t}{\nu_t \! - \! 2} \Big) \, . \label{eq:CE} \\
    &= \text{constants} \ + \label{eq:CE_short} \\
    &\quad \frac{1}{2v_{*}} \Big( (\mu_k^{\intercal} \phi_t - m_{*})^2 + \frac{\beta_k}{\alpha_k}\big(\phi_t^{\intercal} \Lambda_k^{-1} \phi_t + 1\big)\frac{2\alpha_k}{2\alpha_k-2} \Big) . \nonumber
\end{align}  
The mutual information is with respect to the joint distribution in Eq.~\ref{eq:p-future}. Marginalizing $p(y_t,\theta, \tau \given u_t)$ over $y_t$ produces $p(\theta, \tau \given \mathcal{D}_k)$ and marginalizing it over $\theta, \tau$ produces $p(y_t \given u_t)$ (Lemma~\ref{eq:lemma2}). As such, we can factorize the mutual information into three differential entropies \cite{mackay2003information}:
\begin{align} \label{eq:MI}
    &\mathbb{E}_{p(y_t, \theta, \tau \given u_t)} \Big[\ln \frac{p(y_t, \theta, \tau \given u_t)}{p(\theta, \tau \given \mathcal{D}_k) p(y_t \given u_t)} \Big] \\
    & \qquad = - H[\, y_t,\theta,\tau \given u_t ] + H[\, \theta,\tau] +  H[y_t \given u_t]  \, . \nonumber
\end{align}
For the first entropy, note that $p(y_t,\theta, \tau \given u_t)$ consists of Eq.~\ref{eq:postpred_pytau} with a Gamma distribution:
\begin{align}
    p(y_t,\theta, \tau | u_t) \! = \! \mathcal{N}\big( \! \begin{bmatrix}\theta \\ y_t\end{bmatrix} \! | \! \begin{bmatrix} \mu_k \\ \mu_k^{\intercal}\phi_t \end{bmatrix} \! ,  (\tau \bar{\Lambda}_t)^{\-1}\big) \Gamma(\tau | \alpha_k, \beta_k)
\end{align}
where the precision matrix is
\begin{align}
    \bar{\Lambda}_t = \begin{bmatrix} \Lambda_k^{-1} & \Lambda_k^{-1}\phi_t \\ \phi_t^{\intercal} \Lambda_k^{-1} & \phi_t^{\intercal}\Lambda_k^{-1}\phi_t + 1 \end{bmatrix}^{-1} \, .
\end{align}
The differential entropy of a multivariate Normal - univariate Gamma distribution is \cite[ID: P238]{Soch}:
\begin{align}
    H[&\, y_t, \theta, \tau \given u_t ]
     = \frac{D \! + \! 1}{2}\ln(2\pi) + \frac{D \! + \! 1}{2} - \frac{1}{2} \ln |\bar{\Lambda}_t| \label{eq:entropy3} \\
    & + \alpha_k  +  \ln \Gamma(\alpha_k)  -  \frac{D \! - \! 1 \! + \!  2\alpha_k}{2}\psi(\alpha_k)  +  \frac{D \! - \! 1}{2}\ln\beta_k \, , \nonumber 
\end{align}
where $\psi(\cdot)$ refers to a digamma function.
The determinant of the precision matrix can be simplified to:
\begin{align}
    |\bar{\Lambda}_t| \! = \! \big(|\Lambda_k^{\-1}| \, | \phi_t^{\intercal}\Lambda_k^{\-1}\phi_t \! + \! 1 \! - \! \phi_t^{\intercal}\Lambda_k^{\-1}\Lambda_k\Lambda_k^{\-1}\phi_t|\big)^{\-1} \! = \! |\Lambda_k| .
\end{align}
If we plug this result back into Eq.~\ref{eq:entropy3}, then we see that this differential entropy term actually does not depend on $u_t$.
The differential entropy of the parameter posterior distribution (Eq.~\ref{eq:truepost}) evaluates to \cite[ID: P238]{Soch}:
\begin{align}
    H[&\, \theta, \tau] 
    =\frac{D}{2}\ln(2\pi) + \frac{D}{2} - \frac{1}{2}\ln |\Lambda_k| \label{eq:entropy2} \\
    & + \alpha_k +  \ln \Gamma(\alpha_k)  - \frac{D \! - \! 2 \! + \! 2\alpha_k}{2}\psi(\alpha_k) + \frac{D \! - \! 2}{2}\ln\beta_k \, . \nonumber 
\end{align}
This entropy also does not depend on $u_t$.
For $H[y_t |u_t]$, note that a location-scale T-distributed variable is equivalent to a linearly transformed standard student's T-distributed variable. 
Differential entropy is invariant with respect to translation, but follows $H[s y] = H[y]+\ln |s|$ with respect to scaling \cite{mackay2003information}. Thus, we have:
\begin{align}
    &H[ y_t \given u_t] 
    = \frac{2\alpha_k \! + \! 1}{2}\big(\psi(\frac{2\alpha_k \! + \! 1}{2})  -  \psi(\frac{2\alpha_k}{2}) \big)  \label{eq:entropy1} \\
      &\qquad + \ln \sqrt{2\alpha_k} \, B\big(\frac{2\alpha_k}{2},\frac{1}{2}\big) +  \frac{1}{2} \ln \big( \frac{\beta_k}{\alpha_k} (\phi_t^{\intercal} \Lambda_k^{-1} \phi_t + 1) \big) \, , \nonumber
\end{align}
with $B(\cdot)$ the beta function. Only the last term here depends on $u_t$ (through $\phi_t$).
%
%
Plugging Eqs.~\ref{eq:entropy3}, \ref{eq:entropy2} and \ref{eq:entropy1} into Eq.~\ref{eq:MI} and plugging Eqs.~\ref{eq:MI} and \ref{eq:CE_short} into Eq.~\ref{eq:EFE4} completes the proof:
\begin{align}
    &\cJ(u_t) = \text{constants} - \frac{1}{2} \ln \big(\phi_t^{\intercal} \Lambda_k^{-1} \phi_t + 1 \big) \\
    &\ + \frac{1}{2v_{*}} \Big( (\mu_k^{\intercal} \phi_t - m_{*})^2 \! + \! \frac{\beta_k}{\alpha_k}\big(\phi_t^{\intercal} \Lambda_k^{-1} \phi_t \! + \! 1\big)\frac{2\alpha_k}{2\alpha_k-2} \Big) \,  . \nonumber
\end{align}
\end{proof}


\bibliographystyle{IEEEtran}
\bibliography{camready}

\end{document}